# Switching the magnetization of magnetostrictive nanomagnets from single-domain to "non-volatile" vortex states with a surface acoustic wave

Vimal Sampath[1], Noel D'Souza[1], Dhritiman Bhattacharya[1], Gary M. Atkinson[2], Supriyo Bandyopadhyay[2], and Jayasimha Atulasimha[1, 2] *

[1]Department of Mechanical and Nuclear Engineering

[2]Department of Electrical and Computer Engineering

Virginia Commonwealth University, Richmond, VA 23284, USA.

* Corresponding author: jatulasimha@vcu.edu

We report manipulation of the magnetic states of elliptical cobalt magnetostrictive nanomagnets (of nominal dimensions ~ 340 nm × 270 nm × 12 nm) delineated on bulk 128° Y-cut lithium niobate with Surface Acoustic Waves (SAWs) launched from interdigitated electrodes. Isolated nanomagnets that are initially magnetized to a single domain state with magnetization pointing along the major axis of the ellipse are driven into a vortex state by surface acoustic waves that modulate the stress anisotropy of these nanomagnets. The nanomagnets remain in the vortex state until they are reset by a strong magnetic field to the initial single domain state, making the vortex state "non-volatile". This phenomenon is modeled and explained using a micromagnetic framework and could lead to the development of extremely energy efficient magnetization switching methodologies.

Nanomagnetic logic and memory devices have the benefit of non-volatility and internal energy efficiency,[1,2] but it is a challenge to be able to switch the magnetic state of nanomagnets with low dissipation for good external energy efficiency. The usual routes to switching magnetization include the use of electric current-generated magnetic field,[3] spin transfer torque,[4] current-driven domain wall motion,[5] spin orbit torque generated with spin Hall effect,[6,7] spin diffusion from the surface of a topological insulator,[8] voltage controlled magnetic anisotropy[9] and strain generated by applying an electrical voltage to a two-phase multiferroic nanomagnet consisting of a magnetostrictive layer elastically coupled to an underlying piezoelectric layer.[10–18] The energy dissipated in strain-induced switching has been theoretically estimated to be as low as 0.6 atto-Joules,[16,17] making it exceptionally energy-efficient. Switching of magnetostrictive Co[15] and FeGa nanomagnets[18] with feature sizes of 250 – 300 nm with strain generated in an underlying piezoelectric substrate has been reported recently and estimates show that the energy dissipation in ~100 nm scaled structures would have been a mere 4-5 aJ.[18]

Strain can be generated in a two-phase multiferroic nanomagnet by direct application of a voltage (or electric field) across the piezoelectric layer[19,20] using contact pads. However, this would be lithographically challenging in an array of nanomagnets of feature size ~100 nm and pitch 300-500 nm as it would require individual contact pads around each nanomagnet. A global electric field can be used to stress all magnets simultaneously, but this approach has two drawbacks: First, the voltage generated by the field will be very large (resulting in large energy dissipation), and second, this precludes addressing individual nanomagnets *selectively*. Alternatively, a Surface Acoustic Wave (SAW) can be used to stress an array of nanomagnets sequentially as the wave propagates along the array. This has the advantage of allowing sequential stressing (writing bits one at a time in a pipelined manner provided the SAW velocities are sufficiently small) as opposed to simultaneous stressing. More importantly, this reduces the energy dissipation dramatically. There is also no need for lithographic contacts to individual nanomagnets[21] which reduces the fabrication complexity enormously.

The use of SAWs to lower the total energy dissipation in switching of nanomagnets with spin transfer torque has been studied theoretically in the past.[22] Additionally, the periodic switching of magnetization between the hard and the easy axis of 40 μm × 10 μm × 10 nm Co bars sputtered on GaAs[23] and Ni films[24] and excitation of spin wave modes in a (Ga, Mn) As layer by a pico-second strain pulse has been demonstrated.[25] In in-plane magnetized systems, SAWs have been used to drive ferromagnetic resonance in thin Ni films.[26,27] Recent theoretical work has discussed the possibility of complete reversal of magnetization in a nanomagnet subjected to acoustic pulses.[28,29]

In this paper, we demonstrate SAW-based magnetization switching from single domain to vortex state in isolated nanoscale elliptical cobalt nanomagnets that are not dipole coupled to any other nanomagnet. Magnetic Force Microscopy (MFM) is used to characterize the nanomagnets' magnetic state before and after the SAW clocking cycle. The pre-stress state, which is a single domain state, goes into a vortex state upon application of the SAW, and *remains* in the "vortex" state even after the SAW has propagated through and there is no longer any strain in the nanomagnets. The vortex state is therefore non-volatile and a strong magnetic field has to be applied to 'reset' the magnetization of the nanomagnets to its initial single domain magnetic state. We also show that micromagnetic simulations used to study the magnetization dynamics during stress application successfully predict the formation of the non-volatile vortex state when the single domain state is perturbed by the SAW-generated strain.

SAWs are excited and detected by aluminum interdigitated transducers (IDTs) fabricated on a piezoelectric lithium niobate substrate. Elliptical Co nanomagnets are delineated in the delay line of the SAW, as explained in the Methods section and shown in Figures 1d and 1e. The magnetostrictive nanomagnets of nominal dimensions (340 nm × 270 nm × 12 nm) are initially magnetized along the major axis with a large external magnetic field of ~0.2 Tesla (Figure 1a) and characterized by MFM (Figure 2a). The magnetization orientation (single-domain) of these nanomagnets is found to be along the major axis, as expected.

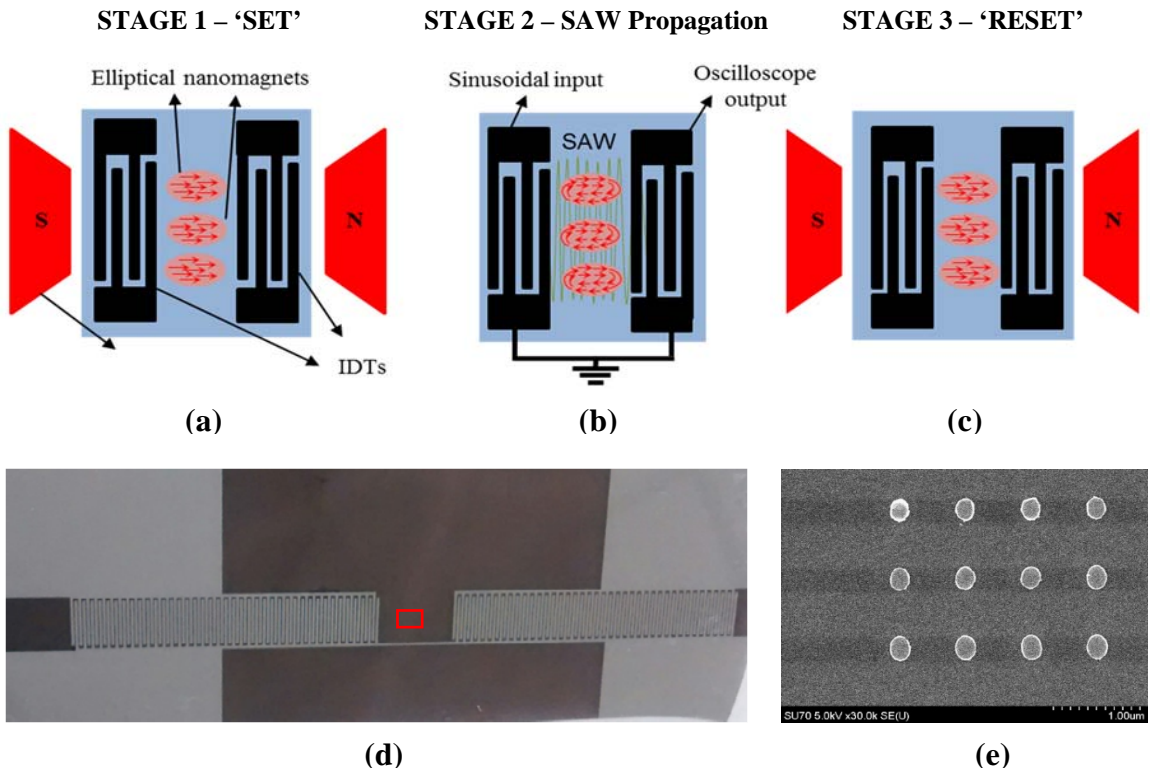

**Figure 1.** (a) Schematic of experimental set-up with initial application of an external magnetic field on the nanomagnets. The arrows indicate the direction of the magnetization state of the nanomagnets. (b) Upon SAW propagation, a mechanical strain is generated and transferred to the nanomagnets which switches the magnetization of the nanomagnets to a 'vortex' state. (c) 'Reset' of the nanomagnets' magnetization by the external magnetic field. (d) SEM micrograph of the lithium niobate substrate with the fabricated IDTs. The red rectangle highlights the region containing the nanomagnets in the delay line. (e) SEM image of the nanomagnets with nominal dimensions of 340 nm × 270 nm × 12 nm.

A SAW is generated and propagated along the surface of the substrate by applying a sinusoidal voltage of 50 $V_{p-p}$ between the IDT's (Figure 1d). This sinusoidal voltage is applied at a frequency of 5 MHz (characteristic frequency of the fabricated IDTs). The relationship between electrostatic surface potential, $\phi$, and applied sinusoidal voltage, $V$, is

$$\phi = \mu(f) \cdot V \qquad (1)$$

Here, $\mu(f)$ is the transmitter response function, which is calculated as 2.016 (see the 'Methods' section). The resulting electrostatic surface potential associated with the SAW in the delay line of Lithium Niobate is 100.8 V, as described in the 'Methods' section. The particle displacement is 0.18 nm per volt of electrostatic potential.[30] The maximum strain generated by this surface acoustic wave in the substrate over a length of 340 nm length is calculated to be 142.16 ppm, as explained in the Supplementary Information. Assuming that this maximum strain is completely transferred to the nanomagnet, the maximum stress generated in the elliptical Co nanomagnets is ~30 MPa. Each nanomagnet experiences cycles of tensile and compressive stress (±30 MPa) along its major axis corresponding to the crest and trough of the SAW. We note that multiple cycles of SAW pass through the nanomagnet, not just a single pulse. Since cobalt has negative magnetostriction, the tensile stress on the nanomagnet results in magnetization rotation towards the minor axis. This is because tensile stress anisotropy shifts the minimum of the nanomagnet's potential energy profile to a location that corresponds to the magnetization being perpendicular to the stress axis. When the tensile stress is removed, the potential energy landscape should revert to its original symmetric double well profile (favoring a magnetization orientation along either direction collinear with the major axis). Therefore, the magnetization should have equal probability of either returning to its original orientation along the major axis or switching its orientation by 180°. Furthermore, when the compressive stress cycle is applied, a magnetization orientation parallel to the major axis is preferred and this will only reinforce the magnetization state that existed at zero stress. Thus, after SAW propagation through the nanomagnets, one would ideally expect the magnetization to be along the major axis, i.e. either point along the original direction or switch by 180°. However, as seen in the MFM image of Figure 2b, after SAW propagation (Stage 2), the magnetization state is no longer of a single-domain nature with orientation parallel to the major axis. Instead, the magnetization goes into a 'vortex' state which persists indefinitely after removal of the SAW.

Here, we note that the MFM tip is magnetized in a direction perpendicular to the substrate, it is sensitive to variations in the perpendicular (out-of-plane) component of the magnetic stray field, with 'bright' ('dark') regions indicating a repulsive (attractive) force between the tip and sample. Thus, there is strong bright and dark contrast along opposite ends of the major axis when the nanomagnet is in a near single domain state with the magnetization pointing along the major axis (Figure 2 a, c). Severe diminishing of this contrast is representative of a continuous magnetization direction variation with low emanating stray field, which demonstrates a vortex state (Figure 2 b). This experimental observation is later rigorously corroborated by detailed micromagnetic simulations. These simulations also explain why the tensile stress spawns the vortex state, why it is stable and why it is preserved when the subsequent compressive stress cycle is applied or as the stress is completely withdrawn. The MFM images in Figure 2b represent the magnetic state of the nanomagnets *after* the SAW has propagated and is not an in-situ visualization.

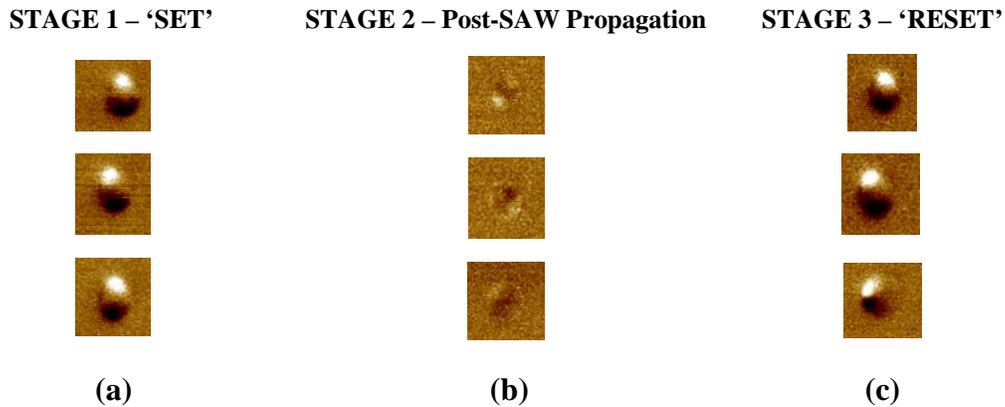

**Figure 2.** MFM images of 3 different Co nanomagnets. (a) In the pre-stress state prior to SAW propagation, the nanomagnets possess a single-domain magnetic state after being initialized by a magnetic field applied along the major axis, (b) In the post-stress state after SAW propagation, the magnetization enters into a stable 'vortex' state. (c) Finally, the nanomagnets are 'reset' by a magnetic field along the same direction as the initial state.

To escape from the vortex state and 'reset' the magnetization of the nanomagnets to the original single domain magnetization state oriented along the major axes, a large external magnetic field of 0.2 Tesla is applied along the major axes of the nanomagnets in the manner shown in Figure 1c. The MFM images of exactly the same nanomagnets after this 'reset' step are shown in Figure 2c. The images clearly show that the single domain pre-stress state of the magnetization has been restored since the images in Figure 2c are nearly identical to those in Figure 2a.

The experimental results are compared against theoretical predictions of magnetization dynamics computed with the MuMax simulation package.[31] The dimensions of the elliptical Co nanomagnet used in our simulations are 340 nm × 270 nm × 12 nm and conform to the nominal dimensions of the experimentally fabricated nanomagnets. The evolution of the magnetization is investigated from its relaxed pre-stress state to a vortex state upon application of one cycle of tensile stress followed by compressive stress. This stress cycle replicates the stress applied on the nanomagnets during SAW propagation.

The discretized cell size used in the MuMax simulations was 4 nm × 4 nm × 4 nm, implemented in the Cartesian coordinate system. Since MuMax has no built-in mechanism to incorporate the effect of stress application, the material's uniaxial magnetocrystalline anisotropy ($K_{u1}$) is used instead since it has the same contribution as stress to the effective field (see Methods section).

The magnetization dynamics are simulated using the Landau Lifshitz Gilbert (LLG) equation:

$$\frac{\delta \vec{m}}{\delta t} = \vec{\tau} = (\frac{\gamma}{1+\alpha^2}) \times (\vec{m} \times \vec{H_{eff}} + \alpha \times (\vec{m} \times (\vec{m} \times \vec{H_{eff}}))) \quad (2)$$

where $m$ is the reduced magnetization ($M/M_s$), $M_s$ is the saturation magnetization, $\gamma$ is the gyromagnetic ratio and $\alpha$ is the Gilbert damping coefficient. The effective magnetic field ($H_{eff}$) is given by

$$\vec{H}_{eff} = \vec{H}_{demag} + \vec{H}_{exchange} + \vec{H}_{stress} \quad (3)$$

where, $H_{demag}$, $H_{exchange}$ and $H_{stress}$ are the demagnetization (or magnetostatic) field, the effective field due to exchange coupling and effective field due to stress anisotropy, respectively. These are evaluated in the MuMax framework as reported by Vansteenkiste et al.[31] For material parameters typical values for cobalt are used: exchange stiffness A = 2.1 × 10$^{-11}$ J/m, saturation magnetization $M_s$ = 1.42 × 10$^6$ A/m, Gilbert damping constant $\alpha$ = 0.01, magnetostrictive coefficient $(^3/_2)(\lambda_s)$ = -50 ppm.

The results of the MuMax micromagnetic simulations for an elliptical Co nanomagnet of dimensions 340 nm × 270 nm × 12 nm subjected to a tensile/compressive stress cycle are illustrated in Figure 3. The initial magnetization state of the nanomagnets is set to the (↑) direction along the major axis. However, when the spins are allowed to relax, the magnetization states settle to the configuration as shown in Figure 3a which shows that not all the spins point along the major axis (as in a perfect single-domain state) and there is some deviation around the edges. Note that the counter-clockwise orientation of the spins in the vortex state is due to the initial conditions applied in our simulation. If the initial magnetization state was such that the pre-stress relaxed state settled to where the spins had a small component in the clockwise direction, this would give rise to a clockwise vortex state.

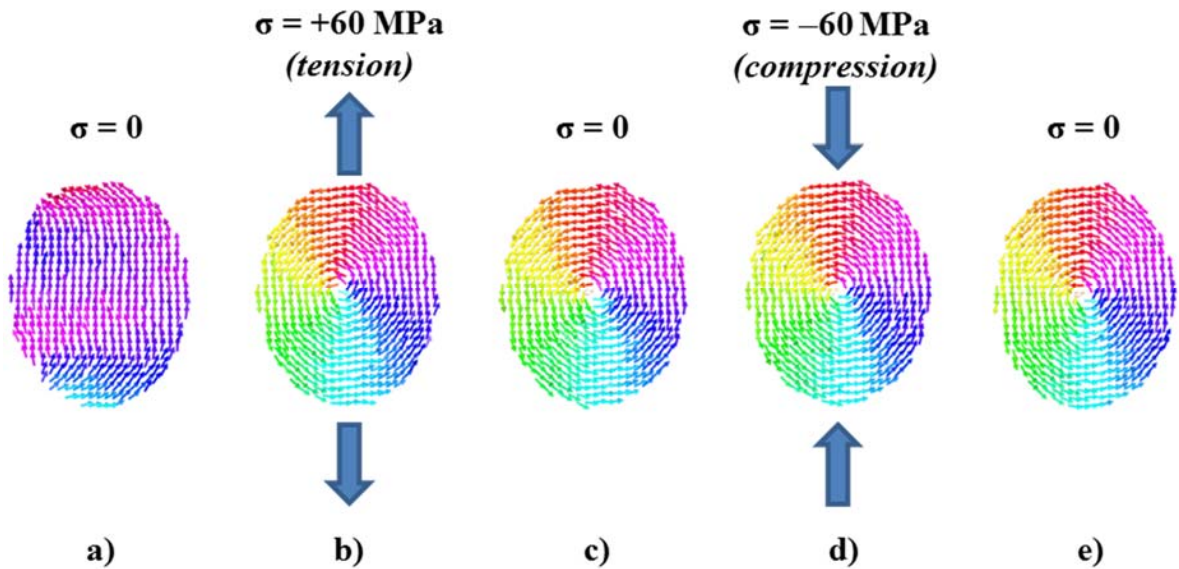

**Figure 3.** Micromagnetic simulations of a nanomagnet with dimensions of 340 nm × 270 nm × 12 nm for the following scenarios: a) Relaxed pre-stress state, b) Tensile stress of +60 MPa, c) Post-stress at 0 MPa, d) Compressive stress of -60 MPa, and e) Post-stress at 0 MPa.

Next, when the sinusoidal stress cycle is applied to the nanomagnet, the magnetic state transforms to a 'vortex' state at a tensile stress of ~ +60 MPa (Figure 3b) and remains in this state as the stress decays down to 0 MPa (Figure 3c). A compressive stress of -60 MPa is subsequently applied to the nanomagnet in a similar manner but it does not nudge the system out of the vortex state (Figure 3d) and finally after stress is removed, the vortex state continues to persist (Figure 3e). This state is non-volatile.

This switching from a single-domain to a vortex state under tensile stress can be explained by analyzing the various energies involved (exchange, demagnetization, stress and total).[32] The total energy calculations show that the vortex state is a local energy minimum and that there exists an energy barrier between the initial relaxed and final vortex states.[32] When the stress anisotropy energy overcomes this barrier, the magnetization enters a vortex state and remains in this state even after the stress is withdrawn. A tensile stress not only drives the magnetization of the nanomagnet to a vortex state, but also causes a slight increase in the magnetization component along the *minor* axis within the vortex (Figure 3b), albeit with the net magnetization remaining zero. Conversely, a compressive stress results in a slight increase in magnetization component along the *major* axis (Figure 3d). Although the visible difference in the tensile (Figure 3b) and compressive (Figure 3d) states is very small, the tendency for the magnetization to align along the major and minor axes, respectively, is expected. This is because the negative magnetostriction of Co causes the easy axis to lie along the long (short) axis when a tensile (compressive) stress is applied. However, due to the presence of a very stable vortex state, the simulations show only a slight deviation from this state at the maximum compressive or tensile stress applied. This shows that the applied stress is clearly not sufficient to drive the magnetization out of the stable vortex state. The magnetization remains in the vortex state even after removal of the stress and an external magnetic field (applied along the major axis) is required to restore the single-domain magnetization state along the easy (major) axis.

While the micromagnetic simulations support the experimental MFM analysis of magnetization switching from single-domain to vortex states, the magnitude of stress required to overcome the energy barrier and enter the vortex state as predicted by the theoretical simulations (60 MPa) is larger than that required in our experiments. This is not entirely unexpected due to the fact that only uniaxial stress is assumed in the simulations, whereas the SAW produces a tensile stress of ~ +30 MPa along the major axis and a compressive stress of ~10 MPa along the minor axis (calculated from the materials' Poisson's ratio values). Therefore, the net stress experienced by the nanomagnet due to the SAW will be ~40 MPa. Furthermore, stress concentrations due to non-uniformities, cracks, etc. in the vicinity of the fabricated nanomagnets could increase the actual stress seen by these nanomagnets, which would explain the remaining discrepancy.

The energy dissipated during SAW propagation for magnetization switching can be determined by calculating the total power generated by the surface acoustic wave. For a 128° Y-cut lithium niobate substrate having a characteristic surface wave velocity, $v_0$ ~ 4000 m/s [30], operating frequency $f$ = 5 MHz and a surface potential $\phi$ = 100.8 V, the power density per unit beam width is 1333.6 W/m, as described in the Methods section. For the purpose of estimating the energy dissipation per nanomagnet, we consider the nanomagnets used in the experiment having lateral dimensions of 340 nm × 270 nm and assume a 2-dimensional array of such magnets can be designed with a center-to-center separation of ~0.5 µm between the nanomagnets (both along the SAW propagation direction and perpendicular to it). With negligible SAW attenuation at low frequencies (less than 0.1% over a length of 1 cm, at 10 MHz[33]), it can be safely assumed that at a frequency of 5 MHz, the SAW wave can clock a ~2 cm long chain of nanomagnets with minimal attenuation. Considering an IDT beam width of 1 cm, a single SAW cycle can clock 800 million nanomagnets and the energy dissipated per nanomagnet for one SAW cycle (tension and compression) of time period 200 ns is ~3.33 fJ. If the clocking frequency is increased a hundred times to ~500 MHz while the power is kept constant (as less stress over smaller time is needed if materials with large magneto-elastic coupling such as Terfenol-D are used) the energy dissipation can be decreased to a mere ~33 aJ per nanomagnet. This would make the SAW based clocking extremely energy efficient without requiring lithographic contacts to each and every nanomagnet.

In summary, we have shown that Surface Acoustic Waves, generated by IDTs fabricated on a piezoelectric lithium niobate substrate, can be utilized to manipulate the magnetization states in elliptical Co nanomagnets. The magnetization switches from its initial single-domain state to a vortex state after SAW stress cycles propagate through the nanomagnets. The vortex states are stable and the magnetization remains in this state until it is 'reset' by an external magnetic field. Furthermore, micromagnetic simulations performed using the MuMax package corroborate the vortex state formation and their being stable under subsequent tensile/compressive stresses. These results lay the foundation for energy efficient switching of nanomagnets with SAW.

## METHODS

**Design and fabrication of the Interdigitated Transducers (IDTs)**

A 128° Y-cut lithium niobate substrate is known to have a characteristic SAW velocity of approximately 4000 m/s.[30] For a SAW frequency of 5 MHz, the SAW wavelength is 800 μm. The Interdigitated transducers (IDTs) are a comb-like arrangement of rectangular aluminum bars of thickness 300 um and gap of 100 μm. The pitch of the IDTs is 400 μm which is exactly half the value of the intended SAW wavelength.

The transmitter response function is $\mu(f)$, which is a function of the frequency of applied voltage, $f$. This transmitter response function is, in turn, a product of the single tap response function $\mu_s(f, \eta)$ and array factor, $H(f)$:

$$\mu(f) = \mu_s(f, \eta) H(f)$$

$$\mu_s(f, \eta) = \mu_s(f_0, \eta) \sin(\pi f/2f_0)$$

$$H(f) = N \sin N\pi [(f-f_0/f_0] / N\pi [(f-f_0/f_0] \qquad (4)$$

The single tap response function varies with frequency, f, and the metallization ratio, η. For an applied frequency of $f = f_0 = 5$ MHz and a metallization ratio of 0.75, $\mu_s(f, \eta) = 0.9K^2$ where $K^2 = 0.056$.[30] When the frequency of applied voltage is equal to the characteristic frequency of the IDTs, the array factor is equal to the pairs of electrodes in the transmitter IDT, N, which is 40 in the current design. Therefore, the transmitter response function, $\mu(f)$, is calculated to be 2.016.

The IDTs are fabricated using conventional photolithography and wet etching processes. Two sets of IDTs are fabricated (as shown in Figure 1). One set is used to launch the SAW and the other is used to sense the propagated SAW. For the purpose of our experiments, the receiver transducer is redundant and is only used to check electrical connections and confirm the propagation of SAW.

**Design and fabrication of the nanomagnets**

Elliptical Co nanomagnets of dimensions ~ 340 nm × 270 nm × 12 nm were fabricated on a 128° Y-cut lithium niobate substrate. Prior to nanomagnet delineation, the substrate was spin-coated with a bi-layer PMMA e-beam resist of different molecular weights in order to obtain a greater degree of undercut: PMMA-495 (diluted 4% V/V in Anisole) followed by PMMA-950 (diluted 4% V/V in Anisole) at a spin rate of 2000 rpm. The resists were baked at 90° C for 5 minutes. Next, electron-beam lithography is performed using a Hitachi SU-70 Scanning Electron Microscope (at an accelerating voltage of 30 kV and 60 pA beam current) with a Nabity NPGS lithography system. Subsequently, the resists were developed in MIBK:IPA (1:3) for 270 seconds followed by a cold IPA rinse. For nanomagnet delineation, a 5 nm thick Ti adhesion layer was first deposited using e-beam evaporation at a base pressure of ~2 × 10$^{-7}$ Torr, followed by the deposition of 12 nm of Co. The liftoff was carried out using Remover PG solution.

**Micromagnetic Modeling: Use of uniaxial anisotropy field to incorporate stress**

Since there is no inherent mechanism of incorporating stress in the micromagnetic software package, MuMax, the material's uniaxial magnetocrystalline anisotropy ($K_1$) is used instead, which is modeled using the following effective field due to the magnetocrystalline anisotropy,

$$\vec{H}_{anis} = \frac{2K_{u1}}{\mu_0 M_{sat}}(\vec{u}\cdot\vec{m})\vec{u} + \frac{4K_{u2}}{\mu_0 M_{sat}}(\vec{u}\cdot\vec{m})^3\vec{u} \tag{5}$$

where $K_{u1}$ and $K_{u2}$ are the first and second order uniaxial anisotropy constants, $M_{sat}$ is the saturation magnetization, $\vec{u}$ and $\vec{m}$ are the unit vectors in the direction of the anisotropy and magnetization, respectively. Assuming $K_{u2} = 0$,

$$\vec{H}_{anis} = \frac{2K_{u1}}{\mu_0 M_{sat}}(\vec{u}\cdot\vec{m})\vec{u} \tag{6}$$

The effective field due to an applied external uniaxial stress, $\sigma$, is

$$\vec{H}_{stress} = \frac{3\lambda_s \sigma}{\mu_0 M_{sat}} (\vec{s} \cdot \vec{m})\vec{s} \tag{7}$$

where $(3/2)\lambda_s$ is the saturation magnetostriction, $\sigma$ is the external stress and $\vec{s}$ is the unit vector in the direction of the applied stress. To simulate the effect of a uniaxial stress applied in the same direction as the uniaxial anisotropy, we equate $\vec{H}_{anis}$ with $\vec{H}_{stress}$ in order to determine the value of $K_{u1}$, as

$$K_{u1} = \frac{3\lambda_s \sigma}{2} \tag{8}$$

**Energy dissipation in nanomagnets due to the surface acoustic wave**

The total power per beam width generated by a surface acoustic wave by application of a voltage to the IDTs is given by[30]

$$\frac{P}{W} = \frac{1}{2} \phi^2 \left(\frac{y_0}{\lambda}\right) \tag{9}$$

where $\phi$ is the surface potential (~100.8 V to generate a stress of 30 MPa), $y_0$ is the admittance of the lithium niobate substrate ($0.21 \times 10^{-3}$ S), $W$ is the beam width of the IDTs, and $\lambda$ is the SAW wavelength ($\lambda = v_0/f$). The surface wave velocity, $v_0 \sim 4000$ m/s,[30] with an operating frequency, $f = 5$ MHz. Therefore, the total power density per unit beam width is 1333.6 W/m.

**ASSOCIATED CONTENT**

**Supporting Information**. The Supporting Information is available free of charge on the ACS Publications website at http://pubs.acs.org.

- Consecutive nanomagnet MFM scans to demonstrate no tip-induced effects on magnetization.
- Calculation of Stress on Co nanomagnets due to SAW

**AUTHOR INFORMATION**

**Corresponding Author**

* Email: jatulasimha@vcu.edu

**Author Contributions**

V.S, N.D., S.B and J.A. played a role in conception of the idea, planning the experiments, discussing the data, analyzing the results and writing the manuscript. V.S. fabricated the SAW devices and nanomagnets and characterized the nanomagnetic switching. N.D. assisted with the MFM analysis, D.B performed the micro-magnetic simulations and assisted with discussion and writing about these simulation results, G.A. assisted with the SAW device design. J.A. also coordinated the overall project.

**Notes**

The authors declare no competing financial interests.

**ACKNOWLEDGEMENTS**

This work is supported by the US National Science Foundation under the SHF-Small grant CCF-1216614, CAREER grant CCF-1253370. We also acknowledge the assistance provided by David Gawalt, Joshua Smak and Catherine Fraher who were involved in the initial design and fabrication of the IDTs as part of their senior design project, as well as Shopan Hafiz and Prof. Umit Ozgur for assisting with the initial operation of the pulse generator. Prof. Umit Ozgur also provided comments on the manuscript. We also acknowledge the use of the CNST facility at NIST, Gaithersburg, Maryland, USA for some of the nanofabrication work.

Supporting Information

# Switching the magnetization of magnetostrictive nanomagnets from single-domain to "non-volatile" vortex states with a surface acoustic wave

*Vimal Sampath[1], Noel D'Souza[1], Dhritiman Bhattacharya[1], Gary M. Atkinson[2], Supriyo Bandyopadhyay[2], and Jayasimha Atulasimha[1, 2] ***

[1]Department of Mechanical and Nuclear Engineering

[2]Department of Electrical and Computer Engineering

Virginia Commonwealth University, Richmond, VA 23284, USA.

* Corresponding author: jatulasimha@vcu.edu

**Consecutive MFM scans to demonstrate no tip-induced effects**

The images shown in Figure S1a are the MFM phase images of the three images shown in Figure 2b of the main paper. These images (Figure S1a) represent consecutive scans along the slow-scan axis of the MFM scan, which demonstrate that the magnetic states of the nanomagnets experience no tip-induced magnetization rotation and are also unaffected by the scan direction.

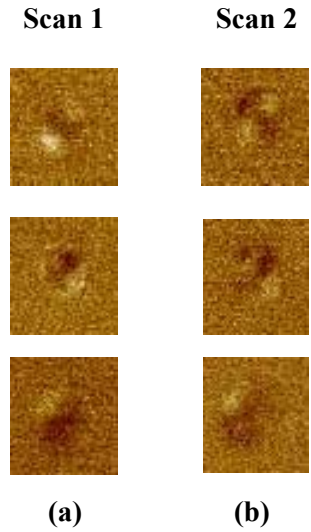

**Figure S1.** (a) MFM phase image of nanomagnets post SAW application (b) MFM phase image with a repeated scan showing no effect of tip.

**Calculation of Stress on Co nanomagnets due to SAW**

The displacement wave can be expressed as $18.1 \sin(2\pi x/\lambda)$ nm, as shown in Figure S2. Here, $\lambda$ is the wavelength of the acoustic wave which is 800 μm. To calculate maximum strain over a length of 340 nm, we need to calculate the displacement at x = 170 nm and x = -170 nm. This is because the strain is maximum around x = 0. The displacement at x = 170 nm is 0.02416 nm and at x = -170 nm is -0.02416 nm. Therefore, the total change in length is 0.04833 nm, and the strain is 0.04833/340 = 142.16 ppm. If we assume the Young's modulus of Co nanomagnets to be equal to the bulk Young's modulus of 209 GPa, then the stress is 29.71 MPa.

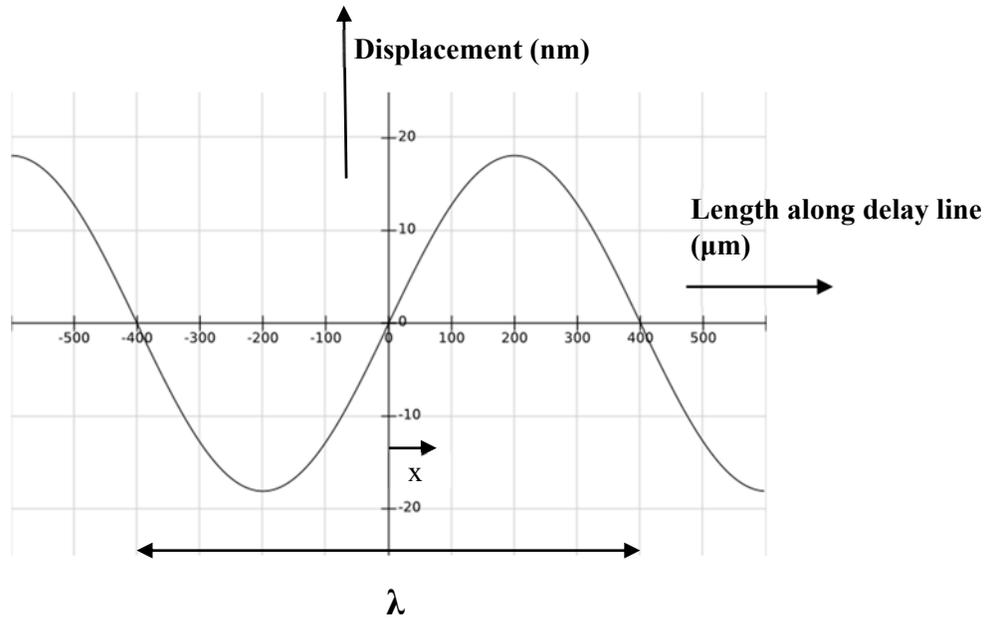

**Figure S2.** The picture showing the displacement wave of points in the delay line of Lithium niobate.